\documentclass[epj,11pt,a4paper]{svjour}

\usepackage[greek,english]{babel}
\usepackage{times,graphics,graphicx,color}
\usepackage{subfigure}
\usepackage{multirow}
\usepackage[bf]{caption} 
\usepackage{amssymb}
\usepackage{cite}
\usepackage{xspace}

\newcommand{\mr}[1]{\ensuremath{\mathrm{#1}}\xspace}
\newcommand{\pt}{\ensuremath{p_\perp}\xspace}
\newcommand{\py}{\textsc{Pythia}\xspace}
\def\professor{\textsc{Professor}\xspace}

\usepackage{scalefnt}
 
\begin{document}


\title{Forward-Backward Correlations and Event Shapes as
  probes of Minimum-Bias Event Properties}
\subtitle{\small{CERN-PH-TH-2011-013, MCNET-11-02, GLAS-PPE/2011-01}}

\author{Kenneth Wraight\inst{1} \and 
Peter Skands\inst{2}}

\institute{Department of Physics and Astronomy, Kelvin Building,
  University of Glasgow, Glasgow G12 8QQ, Scotland, U.K. \and 
European Organization for Nuclear Research, CERN CH-1211, Gen$\grave{e}$ve 23, Switzerland}

\abstract{
\scalefont{1.1}
Measurements of inclusive observables, such as 
  particle multiplicities and momentum spectra, 
  have already delivered important information 
  on soft-inclusive (``minimum-bias'')
  physics at the Large Hadron Collider. In order to gain a more complete
  understanding, however, it is necessary to include
  also observables that probe the \emph{structure} of the studied
  events.  We argue that 
forward-backward (FB) correlations and event-shape observables may be
particulary useful first steps in this respect.
We study the sensitivity of several different types of FB
correlations and two event shape variables --- transverse thrust and
transverse thrust minor ---  to various sources of theoretical
uncertainty: multiple parton interactions, parton showers, colour
(re)connections, and hadronization. The power of each observable to
furnish constraints on Monte Carlo models is illustrated by including
comparisons between several recent, and qualitatively different, 
\py~6 tunes, for $pp$ collisions at $\sqrt{s}=900\,$GeV.   
}

\maketitle

\scalefont{1.06}



\section{Introduction}

Until recently, the measurements used to constrain physics 
models of high energy particle collisions came primarily from
experiments done at the previous generations of accelerators, such as
the SPS, LEP, and the Tevatron. In particular, studies of ``minimum-bias'' and
underlying-event physics have been widely used to constrain the poorly known 
 non-factorizable and non-perturbative aspects of Monte Carlo (MC)
event generators. These generators are, in turn, used ubiquitously
over a continually expanding range of energies and intensities, for
both high- and low-\pt\  processes. 
MC ``tunes'' that rely exclusively on these older data sets are, however, 
becoming outmoded by a new generation of high-energy
experiments, performed at the Large 
Hadron Collider (LHC). The extrapolation of previous results  
to the higher energies and large acceptances of the LHC experiments 
is associated with significant uncertainties, and 
the demands on both theoretical and experimental
precision are becoming ever more stringent. 
The importance of reevaluating the physics models, and of retuning them
\emph{in situ}, has therefore been highlighted in several recent
studies 
\cite{Buttar:2008jx,atlasmc09,Skands:2010ak,Butterworth:2010ym,Abdesselam:2010pt,Deak:2010gk,Aad:2010fh}.

The LHC offers a rich cornucopia of opportunities to test and expand the  
data sets used for MC tuning. Moving from the low-\pt\ results of minimum-bias
to the underlying event in hard processes, and from central rapidities
to ones close to the beam axis, we may test the universality and
 applicability of the modeling over a large dynamical range. 

For
 minimum bias (MB), which is the focus of this paper, there is no
 ``hard scale'', and hence all observables receive large
 non-perturbative corrections. From the point of view of a theoretical
 modeling based on factorization and perturbative QCD, it is difficult to
 say anything meaningful about this data set, except perhaps for the tiny
 fraction of it that includes hard jets. In \py, as in most other
 contemporary MC
 models, the modeling of soft-inclusive physics, is based on the
 concept of Multiple Parton Interactions (MPI). Though these
 corrections go beyond the reach of the standard factorization
 theorems, they are still regarded as essentially perturbative in
 origin; they are dominated by (multiple) $t$-channel gluon exchanges,
 regulated by a non-perturbative dampening at low \pt, and dressed
 with non-perturbative descriptions 
 of the beam remnants and of the hadronization process. 
 We shall not here go into the details
 of the modeling (for a recent review, see \cite{Buckley:2011ms}). It is
 worth noting, however, that the smooth transition
 between soft and hard scattering processes is the main reason the
 \py\ modeling can be used for both minimum-bias and underlying-event
 physics. As mentioned above, this universality is important, since 
 by testing the model in a minimum-bias
 environment here, we may expect to simultaneously improve the description also
 of other physical processes.

Particle multiplicities and transverse momentum spectra are generally
used to give the first 
important constraints on the overall amount of particle production and on its
distribution in $\pt{}$ and $\eta$. Several such studies have already
been published by the LHC collaborations
\cite{Khachatryan:2010xs,Aad:2010rd,Aamodt:2010pp,Aamodt:2010ft,Khachatryan:2010us,Aamodt:2010my,Khachatryan:2010nk,Collaboration:2010ir,ref:MB20}.  
However, as most people familiar
with MC tuning will be aware of, there are often several qualitatively
different ways of mixing the same cocktails. (I.e., several tune parameter
sets may reproduce the same  experimental data.) An important question
therefore concerns the balance between the several different particle
production mechanisms that are available to an open-minded 
model builder: initial- and final-state
radiation, beam remnant breakup, ``hard'' processes vs.\ additional
MPI interactions, final-state interactions, etc. Each production
mechanism generally has some kinematical or dynamical ``signature'' 
that can be used to single it out, if sufficiently differential information
is available.   Tests using several, mutually complementary,  
discriminating observables are therefore to be regarded as essential 
to overcome model degeneracies.

In the context of the present paper, it is especially important to
note that the collinear enhancements characteristic of parton shower
activity tend to produce \emph{local} / \emph{short-range} correlations, i.e., 
the additional particle production caused by the showering dies away rather quickly
with distance to the originating parton. By contrast, the particle
production associated with semi-hard MPI (mini-)jets tends to produce
an enhancement of correlations around $\Delta\phi=\pi$ while soft
particle production from strings stretched between the remnants should
generate long-range correlations in rapidity which, to a first
approximation, should be homogeneous in $\phi$. Thus, the shape of
events and the particle-particle correlations within them, can be used
to gain a handle on the relative strengths of different particle
production mechanisms.

In section \ref{sec:models}, we briefly present the Monte Carlo models
we shall use, and discuss the various sources of particle production
relevant to soft-inclusive physics. A mild selection bias is then
introduced in section \ref{sec:selection}, to mimic a ``minimal''
minimum-bias selection, and the effect of including additional 
$\pt$ cuts is illustrated. In section \ref{sec:current}, we briefly
compare our reference models on 
some of the typical minimum-bias plots, such as charged
multiplicity, $\eta$, and $\pt$ distributions. In section
\ref{sec:new}, we then turn to a more detailed study of
forward-backward correlations, including ones with an explicit $\phi$
dependence, and in section \ref{sec:shapes} we
discuss the transverse thrust and the transverse minor. We give
concluding remarks in section \ref{sec:conclusions}.

\section{Monte Carlo Models and Parameters \label{sec:models}}

For details on the modeling of hadron collisions incorporated in 
general-purpose event generators, we refer the reader to the recent
review  \cite{Buckley:2011ms}. 
Here, we use the \py~6 generator throughout and 
focus on a few main aspects of the modeling that it
will be useful for the reader to be aware of.  

Firstly, the modeling of the diffractive contribution to
soft-inclusive processes in \py~6 is somewhat crude. It 
uses parametrized cross sections to predict the rates of single (SD) and
double (DD) diffractive dissociation differentially in the mass(es) of the
diffracted system(s) \cite{Sjostrand:2006za}. 
Each diffractively excited system is  represented by a single
``string'' of the given mass, which is hadronized according to the 
Lund string fragmentation model
\cite{Andersson:1983ia,Andersson:1998tv}. 
We note that this type of 
diffractive modeling can be characterized as ``soft'' since it does not include a
mechanism for hard, high-mass diffraction, such as diffractive
jet production. We include it, nonetheless, to give an idea of how the
bulk of soft diffractive processes affect our conclusions. We also
note that ``typical'' minimum-bias cuts are designed to reduce the
contamination by diffractive processes, such that our conclusions
should not depend too crucially on the modeling of this component.

The modeling of inelastic, non-diffractive processes is more
sophisticated and is based on a picture of multiple parton
interactions (MPI). In \py~6, there are two basic MPI frameworks available, which we
shall refer to as ``old'' \cite{Sjostrand:1987su} and ``new''
\cite{Sjostrand:2004pf,Sjostrand:2004ef}. (The latter is
similar to the modeling in \py~8.) Briefly stated, 
the main differences between the old and new models are:
\begin{itemize}
\item {\bf Old:} virtuality-ordered parton showers, no showers off the
  additional MPI, and a relatively simple description of the
  fragmentation of the beam remnants in which the baryon number is
  carried by the remnant. 
\item {\bf New:} transverse-momentum-ordered parton showers, including
  showers off the additional MPI, and a more sophisticated treatment
  of the beam remnant, in which ``string junctions''
  \cite{Sjostrand:2002ip} carry the beam baryon number. 
\end{itemize}
In both cases, the fundamental MPI cross sections are derived from a
Sudakov-like unitarization/resummation of perturbative QCD $2\to2$
scattering \cite{Sjostrand:1987su}, normalized to 
the total inelastic non-diffrative cross section, and regulated at low
\pt\ by a smooth dampening factor. The latter is interpreted as being 
due to colour screening, and the dampening scale, $p_{\perp 0}$,
represents the main tunable parameter of the model. Two other significant
parameters are the assumed transverse shape of the proton (lumpy or
smooth), and the strength of colour reconnections (CR) in the final
state, cf.~\cite{Buckley:2011ms}. 

\subsection{PYTHIA Tunes}

We shall consider a small selection of recent tunes that use the ``old'' and 
``new'' models, as follows. 
Field's Tune DW \cite{Albrow:2006rt} 
is currently the ``preferred'' Tevatron tune and has been
extensively tested there. Perugia 0 \cite{Skands:2010ak} 
also attempts to give a good fit to Tevatron data, but uses the
new model and  incorporates an updated set of fragmentation
parameters tuned to LEP data  by the
\professor collaboration \cite{Buckley:2009bj}. That 
collaboration's own tunes of the old and new model are called Q20 and
PT0, respectively \cite{Buckley:2009bj}. Finally, we also include a tune called ACR
\cite{Skands:2007zg}, which 
represents a hybrid between the two models; it uses the basic shower
and MPI framework from the old 
model with the colour-reconnection (CR) model of the new one. It is included
to make it possible to isolate whether specific features are due only
to the CR model or not. All tunes were run with \py\ version
6.4.21. Table \ref{tab:tunes} show the tunes used along with their
three-digit codes in the \py\ subroutine
\texttt{PYTUNE}\footnote{Note: these tunes can also be activated by
  setting the parameter \texttt{MSTP(5)} to the relevant
  \texttt{PYTUNE} code.},  

\begin{table}[!ht]
\begin{tabular}{|l|c|c|c|c|c|}
\hline
Parameter           & DW       & ACR      & Q20      & P0       & PT0      \\
\hline
\texttt{PYTUNE}     & 103      & 107      & 129      & 320      & 329      \\
\hline
\end{tabular}
\centering
\caption{\py\ tunes and corresponding three-digit \texttt{PYTUNE} codes.}
\label{tab:tunes}
\end{table}

\subsection{Sub-Process Samples}

\py\ includes four distinct processes in its simulation of
soft-inclusive physics:  elastic
scattering, single diffractive dissocation, double  
diffractive dissociation, and inelastic non-diffractive (low-\pt) 
interactions. The sum of these
contributions is the total hadron-hadron cross section. 

Elastic scattering occurs when the colliding protons
interact without either of the beam hadrons breaking up. We shall not consider
this source further, since it does not produce any particles at
central rapidities.

In single diffractive dissociation (SD), one of the incoming
hadrons breaks up, and the other does not. In this  
situation, a spread of low-\pt particles is expected from the
disintegrated system over a limited  
rapidity region near the dissociated hadron, 
while the undissociated one continues with a modified momentum.  

Double  
diffractive dissociation (DD) 
involves the break-up of both beam particles. Here, both 
systems generate significant low-\pt particle deposits from
disintegration over rapidity, typically with a gap between them.  

Low-\pt or non-diffractive interaction involves partonic scattering
processes, all the way from soft to hard, with the latter mapping 
smoothly onto the dijet tail. In this case  
particle production is more localized, with higher-\pt
constituents and the possibility --- switched on by default --- of additional
perturbative activity 
such as parton showers and multiple parton interactions.

For all tunes, we start from an inclusive sample composed of the three
inelastic process types, distributed according to their relative cross
sections, which are fixed by \py's default parametrizations
\cite{ref:Pythia_man}. Since the description of the diffractive
components is quite simple,  it would not make much sense to
attempt to isolate individual contributions to the particle production
within the two diffractive samples. The particle production in the 
low-\pt sample, however, receives contributions from several
different algorithmic components which we may label as: ``hard'' scattering, parton showers, MPI, and
remnant fragmentation, each with its own distinct behaviour. 

In order to isolate what happens as each component is ``switched on'',
we consider four different variants of the low-\pt sample:
\begin{enumerate}
\item {\bf low-$\pt$:} Everything on, corresponding to the most physical
  description of the low-\pt\ sample.
\item {\bf HARD:} No parton showers, no MPI. I.e., a single partonic
  $2\to2$ interaction, with no parton showers. 
\item {\bf RAD:} No MPI. I.e., a
  single partonic $2\to 2$ interaction, 
  with parton showers.
\item {\bf MPI:} No parton showers. I.e, multiple parton-parton
  interactions, without showers.
\end{enumerate}
Note that all variants are passed through the string fragmentation model in
order to give final-state hadrons.

For each sample (and for each variation of the low-\pt\ one), 
100,000 $pp$ collisions were generated at $\sqrt{s} = 900\,$GeV. This
sample size is sufficient to overcome statistical fluctuations for the  
measurements of interest, and more than the data set in
\cite{ref:UA5_b}. Obviously,  larger
sample sizes would still be of interest, to study the tails of
distributions, but we shall here mostly be concerned with the bulk of
the physics. 

\section{Selection Procedure}
\label{sec:selection}

Our starting point is a very loose selection, requiring at least
\emph{some} activity in a rather broad central region consistent with
the charged-particle acceptance region of the UA5 experiment. 
Thus, only stable charged particles within a pseudo-rapidity range of
$\eta=(-5.,5.)$ are selected. Although the central
trackers of the ATLAS and CMS experiments 
only extend to pseudorapidities of $\pm2.5$, we note that the Forward
Multiplicity Detector (FMD) \cite{Christensen:2007yc} in the ALICE
experimented could be used to extend charged-particle measurements of
the types we consider to include pseudorapidities up to $\eta 
\sim 5$. By stable charged particles, we mean all charged particles with 
proper lifetimes ${c\tau} > 10\,\mathrm{mm}$ (thus, e.g., $\Lambda$ and
$K$ hadrons are stable). By default, we do not apply any cuts in
\pt\ unless explicitly stated otherwise. Selected events must have
at least one charged particle in the $\eta$-range. 

Table \ref{tab:selection} shows the percentages of selected events
with this mild requirement. For elastic events, the corresponding percentage would be
zero, as the scattered protons continue on ``down the beam-pipe'', 
outside the range of selection. Of the included sub-processes, a
significant fraction of SD events are rejected already at this point, 
as are about half that fraction of the events labeled DD. These
correspond to events where the fragments of the dissociated proton(s) 
were concentrated at high rapidities, beyond the acceptance
region. The events in the DD sample are less likely to be rejected, 
since they have two ``chances'' to produce particles in the central
region. The low-\pt\ ones are selected with approximately 100\%
efficiency\footnote{For completeness, we note that a few of the
  generated low-\pt events do fail, below the per
mille level, having produced no or only neutral particles in the
central region.}. The ``mix'' sample contains the sum of all three
subprocesses, weighted according to their respective cross sections as given by
\py. The particle production samples (hard  
process only, radiative, MPI and combined i.e. low-\pt) have
similar selection rates as they all include a central `hard' interaction. 

\begin{figure}[!ht]
  \centering
    \includegraphics[width=0.50\textwidth]{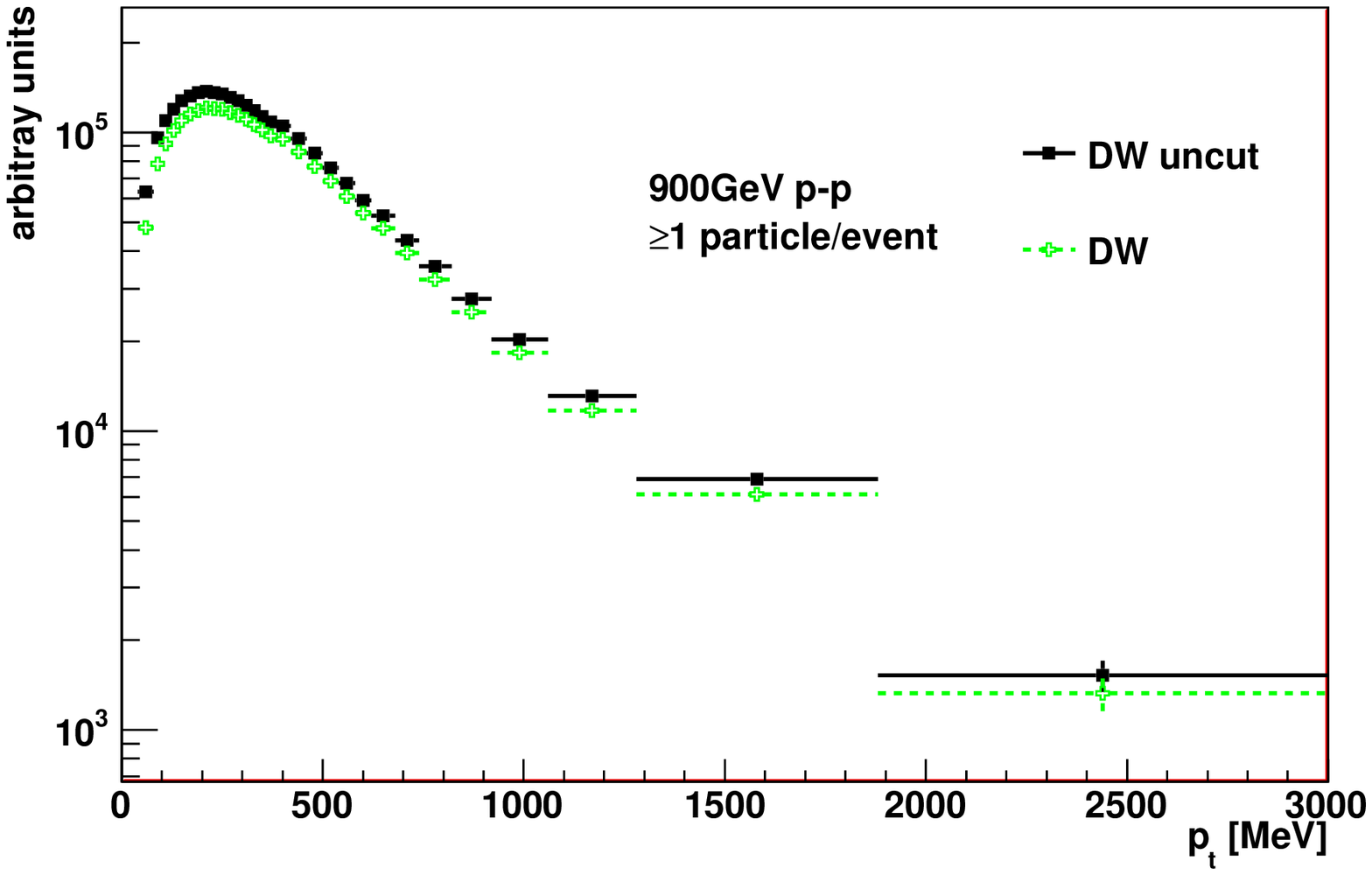}
    \includegraphics[width=0.50\textwidth]{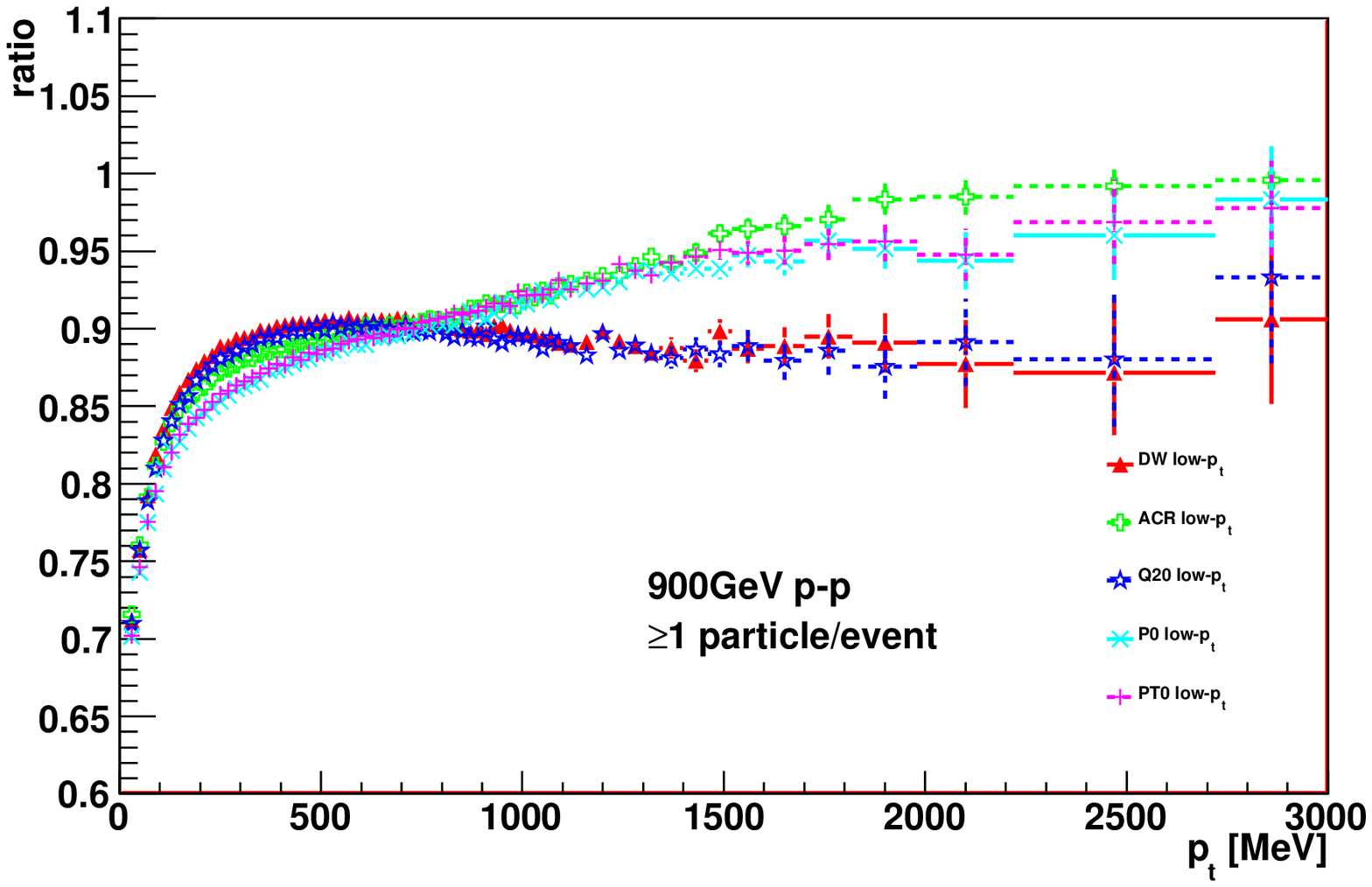}
\caption{Above: Charged particle \pt distribution for
  low-\pt sample of DW tune, with (green) and without (black) $\eta$
  selection cut. Below: Ratio of selected to unselected tracks for the
  low-\pt sample of each tune.}
  \label{fig:selection}
\end{figure}

An illustration of the effect of the event selection on the inclusive
\pt\ distribution is shown in figure
\ref{fig:selection}, for the low-\pt sample. 
The top pane shows the \pt\ distribution of 
all generated tracks in black and of 
the selected ones in green, for the DW tune. The main effect is a
reduction in the total number of accepted tracks by 10-15\%. A
secondary effect, however, is a model-dependent hardening of the
spectrum. We highlight this in the lower
pane, which shows the ratio of selected to unselected tracks for each
of the tunes. We observe that the DW and Q20
tunes of the ``old'' model exhibit an approximately constant value of
this ratio, indicating that the shape of the \pt\ spectrum is not
greatly different at high rapidities than in the central region. By
contrast, the tunes of the ``new'' model (P0 and PT0) exhibit a
noticeable shape, indicating that for those models, the spectrum of
the unselected high-rapidity tracks is systematically softer than in
the central region.
Also, since the hybrid ACR tune follows the ``new''
tunes here, we may interpret this behaviour as being related to the
CR model used in the new tunes. 

We note that these differences would
translate directly to uncertainties for any purely model-based
extrapolation beyond the central region. For the remainder of this
report, however, we shall be concerned only with the central region
itself. 

\begin{table}[t]
\resizebox{8.5cm}{!}{
\begin{tabular}{|l|c|c|c|c|c|c|c|c|}
\hline
                    & mix    & SD  & DD  & low-$p_{t}$  & HARD   & RAD    & MPI    \\
\hline
DW                  & 73\%  & 70\%   & 86\%   & 100\%        & 100\%  & 100\%  & 100\%  \\
ACR                 & 73\%  & 70\%   & 86\%   & 100\%        & 100\%  & 100\%  & 100\%  \\
Q20                 & 73\%  & 69\%   & 86\%   & 100\%        & 100\%  & 100\%  & 100\%  \\
P0                  & 73\%  & 68\%   & 85\%   & 100\%        & 100\%  & 100\%  & 100\%  \\
PT0                 & 73\%  & 68\%   & 85\%   & 100\%        & 100\%  & 100\%  & 100\%  \\
\hline
\end{tabular}}
\centering
\caption{Selection efficiencies for various Pythia sub-processes and particle production mechanisms for each tune. Each sample consisted of 100,000 generated events. }
\label{tab:selection}
\end{table}

\section{Inclusive Distributions \label{sec:current}}

Current LHC studies of minimum-bias events, e.g. \cite{Khachatryan:2010xs,Aad:2010rd,Aamodt:2010pp,Aamodt:2010ft,Khachatryan:2010us,Aamodt:2010my,Khachatryan:2010nk,Collaboration:2010ir,ref:MB20}, have
focused mainly on the ``basic four'' charged-particle distributions:  
$P(n_{ch})$, $dn_{ch}/d\eta$, $dn_{ch}/dp_{t}$, and 
$\left<p_{t}\right>$ 
vs.\ $n_{ch}$. In this section, we comment briefly on these and 
illustrate the behaviour of our chosen set of Monte Carlo tunes for 
later reference. More comments on these distributions can be found, e.g., in 
\cite{Skands:2010ak,Buckley:2011ms}. 

\subsection{Multiplicity}

In table \ref{tab:mult_procs}, we compare the average charged particle multiplicity, 
$\left<n_{ch}\right>$, between tunes, for the three different inelastic samples, as 
well as for their cross-section-based mixture. At this point, we subject 
the samples only to the selection requirement $|\eta|\le 5$ mentioned 
above. Table \ref{tab:mult_parts} contains an equivalent comparison, 
for the different physics variations of the low-\pt
sample. To help illustrate the overall spread in predictions, we also 
quote a ``range'' of variation, at the bottom of each table, 
which is defined as the highest average multiplicity of the tunes minus the lowest,  
normalized to the lowest multiplicity, i.e. 
\begin{equation}
\mbox{range} =
(\left<n_{ch}\right>_{\mr{max}}-\left<n_{ch}\right>_{\mr{min}})/\left<n_{ch}\right>_{\mr{min}}~.\label{eq:range}
\end{equation} 

The range of $\left<n_{ch}\right>$ predicted within $|\eta|\le 5$ 
varies by 10--20\% between the different tunes for each 
of the inelastic sub-processes in table \ref{tab:mult_procs}. For the 
diffractive processes, there is no parton showering and no MPI. The 
considerable differences between models are therefore  
solely generated by the different tunings of the 
hadronization model. Since in particular Q20 and PT0 were tuned to 
exactly the same LEP data by exactly the same tuning program 
(\professor), the 
difference between them here highlights the need for in situ 
constraints on the non-perturbative fragmentation parameters. It also
indicates the need for a Monte Carlo modeling of diffraction that would be more 
theoretically consistent with the treatment at LEP, where the 
fragmentation parameters are dependent, e.g. on the perturbative 
parton shower cutoff. For the time 
being, we conclude that the fragmentation tuning of the new model 
used by PT0 and P0 produces fewer particles in and of itself 
than that of the old models. This appears to be consistent with the 
\pt\ distributions shown in \ref{sec:selection}; fewer particles 
produced from the same energy of collision will result in a larger 
proportion of high-\pt particles in selected events. 
Though not the main focus of this report, 
this would clearly be worth a more detailed analysis especially in the context of 
diffractive studies. 
 
\begin{table}[t]
\begin{tabular}{|l|c|c|c|c|}
\hline
Tune                      &  SD       & DD    & low-\pt   & mix     \\
\hline 
DW                        &  10.1         & 12.0       & 38.6          & 23.5    \\
ACR                       &  10.1         & 12.0       & 38.4          & 23.5    \\
Q20                       &  10.3         & 12.1       & 38.3          & 23.4    \\
P0                        &  9.1          & 10.8       & 35.4          & 21.7    \\
PT0                       &  9.1          & 10.8       & 36.8          & 22.5    \\
\hline 
range in $\left<n_{ch}\right>$ (\%)  &  14.1         & 12.8       & 9.0           & 8.5  \\
\hline
\end{tabular}
\centering
\caption{Average charged particle multiplicity for the various minimum-bias sub-processes and tunes.}
\label{tab:mult_procs}
\end{table}

\begin{table}[t]
\begin{tabular}{|l|c|c|c|c|c|}
\hline
Tune                     & HARD      & RAD      & MPI     & All on    \\
\hline
DW                       & 28.0           & 32.9                & 33.5           & 38.6    \\
ACR                      & -              & 31.6                & 36.2           & 38.4    \\
Q20                      & 29.1           & 30.9                & 36.3           & 38.3    \\
P0                       & 23.8           & 26.0                & 29.6           & 35.4    \\
PT0                      & 24.1           & 26.0                & 31.1           & 36.8    \\

\hline 
range in $\left<n_{ch}\right>$(\%)  & 22.4           & 26.6                & 22.4           & 9.0     \\
\hline
\end{tabular}
\centering
\caption{Average charged particle number for each particle production
  mechanism. Note: the column labeled ``All on'' is identical to the
  ``low-\pt'' one in table \ref{tab:mult_procs}. }
\label{tab:mult_parts}
\end{table}

\subsection{Track \pt }

\begin{figure}[t]
  \centering
  \includegraphics[width=0.50\textwidth]{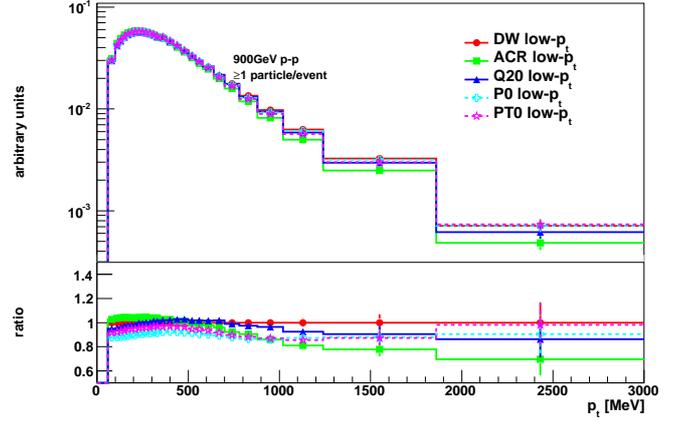}
  \caption{Selected charged particle logarithmic \pt
    distribution for low-\pt sub-sample \py  tunes. Lower pane:
    ratio to the DW distribution.}
  \label{fig:pt_tracks}
\end{figure}

Figure \ref{fig:pt_tracks} shows the \pt\ distributions for the
low-\pt\ sample of each tune, with the range of variation of the
average $\left<\pt\right>$ spanned by the tunes for all the
sub-samples given in table
\ref{tab:pt_ranges}, with the range defined as in eq.~(\ref{eq:range}). 

As before, we note that the tunes exhibit differences of the order
10--20\%. In the low region ($\pt<1\,\mr{GeV}$) the old shower tunes
DW, ACR and Q20 lie above the new shower models P0 and PT0. This
reverses in the region above $2\,\mr{GeV}$. Due to limited statistics,
we do not plot the tail of very high-\pt charged particles  here, 
but note that the trend of the new models to generate harder \pt
tails is illustrated in \cite{Skands:2010ak}. 

\begin{table}[t]
\resizebox{8.5cm}{!}{
\begin{tabular}{|l|c|c|c|c|c|c|c|}
\hline
                        & mix  & SD   & DD   & low-$p_{t}$   & HARD   & RAD   & MPI   \\
\hline
range in $\left<p_{t}\right>$ (\%) & 5.7   & 3.1      & 3.5      & 6.1           & 10.5   & 6.1   & 10.5 \\
\hline
\end{tabular}
}
\centering
\caption{Range of mean $p_{t}$ values for various Pythia sub-processes and particle production mechanisms for each tune.}
\label{tab:pt_ranges}
\end{table}

\subsection{Track $\eta$ }

\begin{figure}[b]
  \centering
  \includegraphics[width=0.50\textwidth]{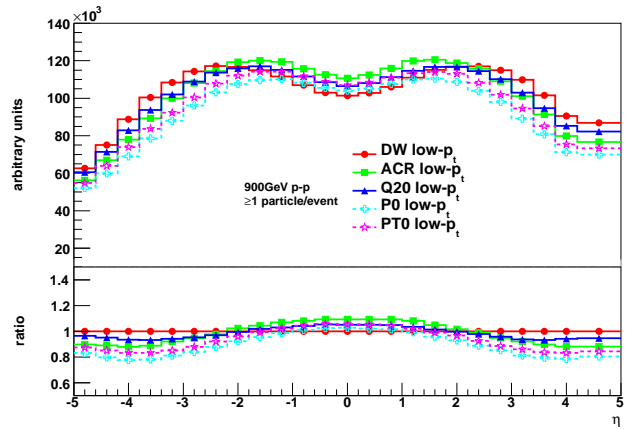}
  \caption{Selected charged particle logarithmic $\eta$ distribution
    for low-\pt\ sub-sample \py tunes. Lower pane: ratio to the DW distribution.}
  \label{fig:eta_tracks}
\end{figure}

Figure \ref{fig:eta_tracks} shows the $\eta$ distributions for the
selected models. Differences are again at the 10--20\%
level. Especially in the context of this distribution, such
differences have sometimes been represented as ``large''. Let us
recall, however, that the \py\ modeling is rooted in
perturbative QCD, that we are here dealing with processes which have
no hard scale, and that the number of charged particles is not an
infrared safe observable. In this light, while we might still hope to
constrain the modeling better, we nonetheless wish to point out that
it is, in our opinion, grossly misleading to characterize 
order 10\% differences 
as large.

Indeed, the small differences between tunes are highlighted by the
zero-suppressed Y-axis in the plot. Thus, while there is clearly
some sensitivity to central vs.\ forward production mechanisms in this
distribution, its ability to discriminate between models is still
limited. Agreement between each tune 
is generally good, especially in the most easily observable 
region, $|\eta|<2.5$. We conclude that additional, linearly independent, 
information on the structure of events in $\eta$, could provide
valuable additional constraints. 

\section{Forward-Backward Correlations \label{sec:new}}

We come now to the main part of this report, in which we study
several types of forward-backward correlations, $b$, for 
different production mechanisms, cuts, and correlation regions.

The purpose of these distributions is to enhance the discriminating power 
between models, and to reveal their properties more clearly, 
as compared to what can be achieved with 
the list of observables discussed in section
\ref{sec:current}. In particular, the collinear singularity structure
of bremsstrahlung corrections in perturbative QCD causes initial- and
final-state shower activity to 
generate strong but primarily short-range 
correlations, spanning at most a few rapidity units, 
whereas coloured exchanges between the beam hadrons (e.g.,
MPI) can generate correlations that are weaker but which 
span the entire rapidity range
between the remnants. Thus, the shapes and  
normalizations of the $b$ distributions contain valuable information
on the relative dominance of different particle production mechanisms,
information which we argue is linearly independent from that contained
in the current ``standard'' distributions. 

This section is divided as follows: 
we first consider a standard inclusive ``minimum-bias'' $b$
correlation in section \ref{sec:b-inclusive}, illustrating how it is
affected by different choices of bin size and by \pt\ cuts; 
in section \ref{sec:b-mechanisms}, we illustrate the sensitivity of this
correlation to different particle production mechanisms, using the
HARD, RAD, and MPI samples defined above, and to 
contamination by diffractive processes (SD, DD). 
In this way, we gain a map of how different cuts and different process
mixtures affect the correlations, that we hope will be 
useful for future reference. 
We shall seek to extract further information by defining also a
set of $b$ correlations 
that are sensitive to the azimuthal structure of the events, which
will be the focus of section \ref{sec:b-twisted}. 
We shall refer to these latter observables, which are essentially
binned double-differential $\eta$-$\phi$ correlations, as ``twisted''
$b$ correlations. 

\subsection{Inclusive b Correlation \label{sec:b-inclusive}}

The standard $b$ correlation is defined as:

\begin{equation}
  \label{eqn:b-correlation}
  \centering
  b = \frac{\sigma(n_{b},n_{f})}{\sigma(n_{b}) \sigma(n_{f})} = \frac{\left<n_{b}n_{f}\right> - \left<n_{f}\right>^{2}}{\left<n_{f}^{2}\right> - \left<n_{f}\right>^{2}}~,
\end{equation}
where $n_{f}$ ($n_{b}$) is the activity in a specific forward (backward)
region of the detector. ``Activity'' can be measured by a number of
observables in the detector, e.g., energy, charged particle
multiplicity (inclusively or above a given \pt\ threshold), momentum sum,
etc. Here, we shall focus on the charged-particle multiplicity, as has
also been done in most previous studies, though we emphasize that, e.g.,
calorimetric energy sums, could also be interesting to 
explore (see, e.g., \cite{Skands:2010ak}).

The ``forward'' and ``backward'' regions are defined by bins of a
specific size in $\eta$ --- typically chosen to be between 0.1 and 1.0 unit
wide in $\eta$ --- which are separated by some variable distance
and arranged symmetrically around a midpoint which is
usually taken to be the centre of the detector, $\eta_c = 0$,
corresponding to the CM of the colliding hadrons. Although we shall
not do so here, we note that correlations between the central and
forward region are also of interest and can be probed, for example, 
by fixing one bin in the central region and letting the other slide into the forward
or backward region, corresponding to choosing $\eta_c \ne 0$. A study somewhat
 along these latter lines has been performed by UA5 \cite{ref:UA5_b}
 and could also be interesting to maximize usage of the asymmetric
 coverage of the ALICE FMD.

The optimum bin size to use in eqn.~(\ref{eqn:b-correlation}) is a
function of  statistics 
and of the $\eta$-range observed. If the bin size is too small, genuine
correlations will be washed out by statistical  
fluctuations. With too large a bin size, the resolving power of the
correlation over the limited $\eta$-range will be  
lost. 

\begin{figure}[t]
  \centering
    \includegraphics[width=0.5\textwidth]{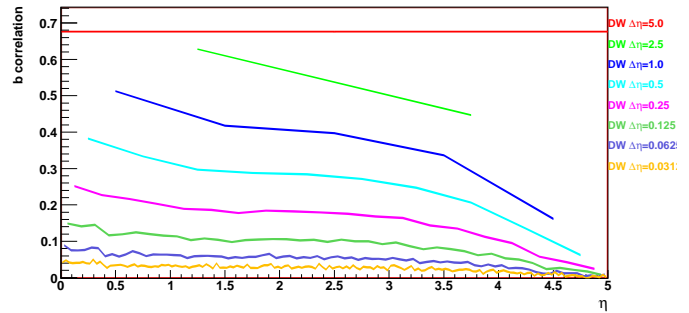}
\caption{$b$ correlation for selected events with various pseudo-rapidity bin sizes, $\Delta\eta$. The single-point correlation for a bin size of $\Delta\eta$=5.0 (red) is shown covering the whole $\eta$-region. }
  \label{fig:bin_size}
\end{figure} 
Figure \ref{fig:bin_size} contains 
a comparison of the $b$ correlation vs.\ $\eta$ (specifically the
$\eta$ value of the centre of the forward bin, with the backward one
located at $-\eta$) for varying bin sizes 
from 0.03 to 5 units wide, without imposing any \pt\ cuts at this
point. 
Obviously, the largest sizes are too coarse
to discern much structure in the correlation distribution. 
Mid-range bin sizes, $\Delta\eta$=1.0, 0.5 and 0.25 exhibit 
best the trends over the $\eta$-range; for this particular model (DW), 
a high correlation at low $\eta$ can be distinguished from a mid-$\eta$ plateau
and a further drop in correlation at high $\eta$. Going to even smaller  
bin sizes, $\Delta\eta$=0.125, 0.0625 and 0.03125, we begin to 
lose the structure in the distribution as statistical fluctuations 
start to dominate. We conclude that a bin size of $\Delta\eta$=0.5 
is reasonable for this study. Note, however, that going to different
CM energies and/or imposing \pt\ cuts could change this conclusion; 
the average accepted multiplicity at each $\eta$ value 
determines the relative size of the statistical fluctuations and
hence affects the optimum bin size.
As a consequence, it is not possible, therefore, to directly compare $b$
distributions taken with different cuts, or which use different-sized
$\Delta\eta$ bins. 

\begin{figure}[t]
  \centering
  \includegraphics[width=0.50\textwidth]{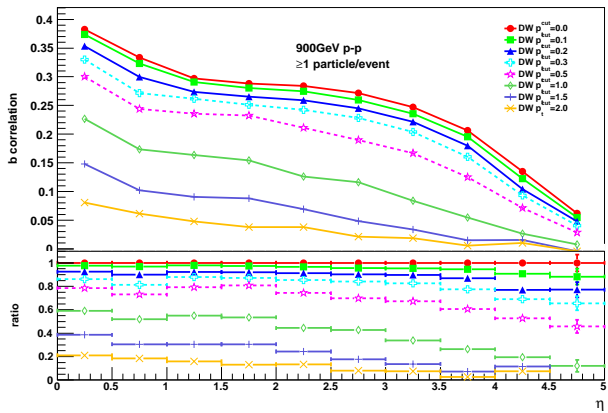}
  \caption{The $b$ correlation distributions for hadron-level charged
    particles for the DW tune with various explicit \pt cuts. Lower
    pane: ratio to the non-cut distribution.}
  \label{fig:DW_pts}
\end{figure}

\begin{figure}[t]
  \centering
  \includegraphics[width=0.50\textwidth]{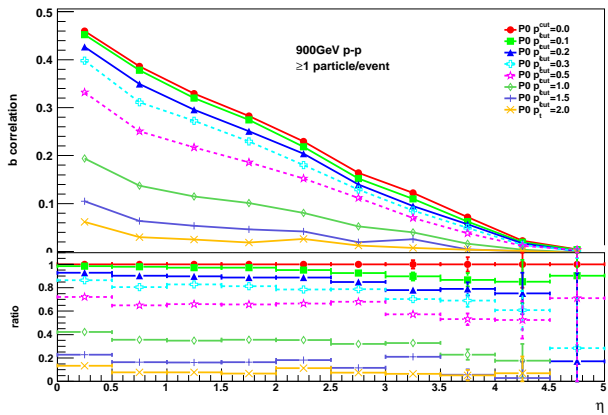}
  \caption{The $b$ correlation distributions for hadron-level charged
    particles for the Perugia 0 tune with various explicit \pt
    cuts. Lower pane: ratio  to the non-cut distribution.}
  \label{fig:P0_pts}
\end{figure}
To illustrate the effect of \pt cuts, 
figs.~\ref{fig:DW_pts} and \ref{fig:P0_pts} show the $b$
correlations subjected to a range of different \pt\ 
cuts for a tune of the old (DW) and new (Perugia 0) model,
respectively. 
(Note that these cuts are applied also at the level of the
event selection, so only events with at least one particle harder
than the given \pt\ cut are included, for each curve.)
One clearly sees the reduction of the 
correlation strengths as fewer tracks make
it into each bin. 
Note that only positive correlations are expected and
plotted. We interpret any small negative correlations 
as arising from statistical fluctuations in poorly populated bins.

\begin{table}[t]
\resizebox{8.5cm}{!}{
\footnotesize
\begin{tabular}{|l|c|c|c|c|c|c|c|c|c|}
\hline
                     & \multicolumn{3}{|c|}{central bin}                    & \multicolumn{3}{|c|}{mid-range bin}                    & \multicolumn{3}{|c|}{extreme bin}   \\
\hline
Tune                 & $b_{0}$  & $b_{0.5}/b_{0}$    & $b_{1.5}/b_{0}$      & $b_{0}$  & $b_{0.5}/b_{0}$    & $b_{1.5}/b_{0}$      & $b_{0}$  & $b_{0.5}/b_{0}$    & $b_{1.5}/b_{0}$       \\
\hline
DW                   & 0.38     & 0.79               & 0.39                 & 0.28     & 0.74               & 0.24                 & 0.06     & 0.46               & -0.08     \\
ACR                  & 0.45     & 0.71               & 0.34                 & 0.28     & 0.68               & 0.30                 & 0.05     & 0.38               & -0.03     \\
Q20                  & 0.43     & 0.74               & 0.29                 & 0.27     & 0.68               & 0.17                 & 0.05     & 0.42               & 0.16      \\
P0                   & 0.46     & 0.72               & 0.23                 & 0.23     & 0.66               & 0.18                 & 0.00     & 0.71               & -0.39     \\
PT0                  & 0.45     & 0.73               & 0.23                 & 0.24     & 0.65               & 0.18                 & 0.01     & 0.15               & -0.05     \\

\hline
\end{tabular}
}
\centering
\caption{Central($\eta$=0-0.5), mid-range ($\eta$=2.5-3) and extreme ($\eta$=4.5-5) correlation values with fraction of correlation remaining after \pt-cut= 0.5GeV and 1.5 GeV for various Pythia tunes at $\sqrt{s}$=900GeV. $b_{0}$ is the correlation value for \pt-cut=0.0GeV, $b_{0.5}$ the value with \pt-cut=0.5GeV and $b_{1.5}$ the value \pt-cut=1.5GeV. }
\label{tab:pt_cuts}
\end{table}
Table \ref{tab:pt_cuts} summarizes the effect of \pt\ cuts by giving
the $b$ correlation values in the central ($|\eta|$=0-0.5), mid-range
($|\eta|$=2.5-3)   
and extreme ($|\eta|$=4.5-5) bins, without any \pt\ cut, 
together with the reduction in the correlation strengths caused by
\pt\ cuts of $500\,$MeV and $1.5\,$GeV. In the   
central $\eta$-region, the effect of the $\pt = 500\,$MeV cut (pink
dashed line in figs.~\ref{fig:DW_pts} and \ref{fig:P0_pts}) 
is to lower the correlation by 20 -- 30\%. When the \pt cut is
increased to 1.5 GeV (dark purple solid line, second from bottom in the
plots), the reduction is much more severe, and more interestingly is
not a simple scaling from the reduction caused by the previous cut. 
Thus, e.g., ACR exhibits the largest reduction with the first cut, but
the second-to-least with the second. Also interestingly, this pattern
changes as one goes from mid-range to the extreme $\eta$ bins. (At the
extreme end, though, the correlations are small, especially with
high-\pt\ cuts, and hence are easily overpowered by statistical
fluctuations.) When comparing correlation 
values surviving the 500 MeV cut between central and extreme
$\eta$-bins, we may conclude that the particle momentum distributions
are heterogeneous across the $\eta$-range, due to the different
parameters between tunes and differing
particle production mechanisms. 
Measurements should therefore by no means be restricted to the most
inclusive definition possible for a given experiment. 

\begin{figure}[t]
  \centering
  \includegraphics[width=0.50\textwidth]{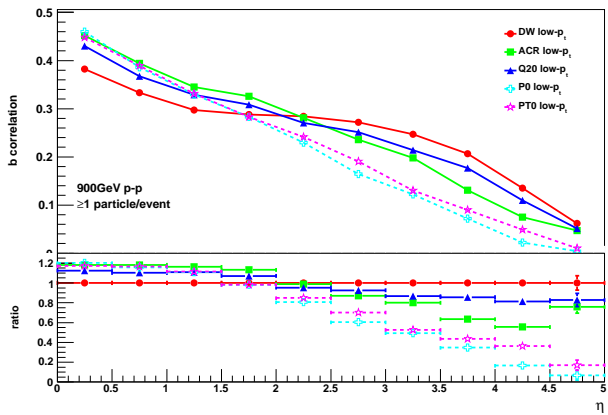}
  \caption{Inclusive $b$ correlation distribution for selected
    low-\pt\ \py\ tunes. Lower pane: Ratio to the DW distribution.  \label{fig:b_corr_all_inc}
}
\end{figure}

As a final summary of this part of the study,
fig.~\ref{fig:b_corr_all_inc} contains a comparison of the
$b$ correlation 
distribution for each of the different tunes, again 
without \pt\ cuts imposed. As was already apparent from
figs.~\ref{fig:DW_pts} and \ref{fig:P0_pts}, the old and new models exhibit
qualitatively different shapes. We interpret this in the following
way: due to the inclusion of showers off
the MPI in the new models, more of their total particle production is driven by
shower activity than what was the case in the old ones, which have
a larger average number of MPI~\cite{Skands:2010ak}. The new models
therefore exhibit stronger 
short-range correlations\footnote{Note that these particular tunes of
  the new model have fewer average charged particles than
  those of the old, cf.\ table \ref{tab:mult_procs}. Due to the
  dilution effect caused by statistical fluctuations, their
  correlation strengths are therefore intrinsically a bit lower than
  if they  had been
made to give the same average multiplicities as their older counterparts. }  
and weaker long-range ones than their older
counterparts, with a crossover point  somewhere around
$|\eta|=1-2$. 

There are further differences in the $b$ correlation distributions
especially within the old model class. DW has the most  
distinctive shape of the old models, with a clear plateau-like
structure at mid-$\eta$ which is not as pronounced for any of the
other models. This is consistent with the $dN/d\eta$ distribution
being higher for this tune for $|\eta|>3$ than for any of the other
models, cf.~fig.~\ref{fig:eta_tracks}. Due to this significant shape
difference, the Q20 distribution, for instance, lies above DW at low
$\eta$, but then drops below it at high $\eta$. It should therefore be
clear that a measurement of the shape of this distribution out to
as high $\eta$ as possible would yield valuable information.

\subsection{Physical Sources of Correlations} 
\label{sec:b-mechanisms}

\begin{figure}[t]
  \centering
  \includegraphics[width=0.50\textwidth]{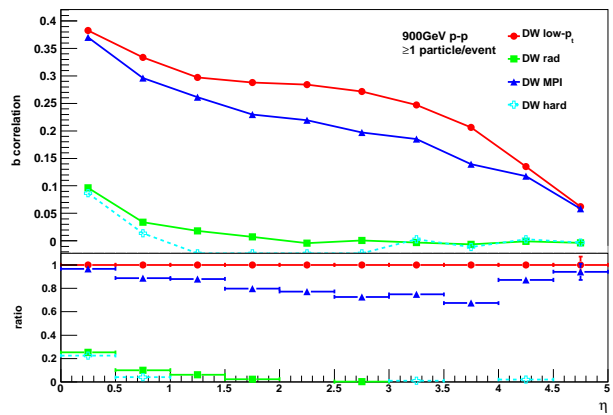}
  \caption{Inclusive $b$ correlation distribution for tune DW particle
    production mechanisms: low-\pt, hard process (HARD), radiative
    production (RAD) and multi-parton interactions (MPI). Lower pane: ratio to the low-\pt distribution.}
  \label{fig:b_corr_DW_parts}
\end{figure}

\begin{figure}[t]
  \centering
  \includegraphics[width=0.50\textwidth]{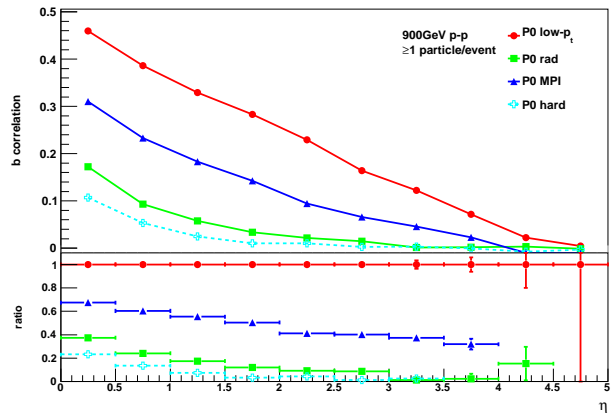}
  \caption{Inclusive $b$ correlation distribution for tune Perugia 0
    particle production mechanisms: low-\pt, hard process (HARD),
    radiative production (RAD) and multi-parton interactions
    (MPI). Lower pane: ratio to the low-\pt distribution.}
  \label{fig:b_corr_P0_parts}
\end{figure}

To investigate the sensitivity to the different sources 
of particle production in more detail, we now turn to the HARD, RAD, and MPI samples,
as compared to the default low-\pt\ sample which has all the physics
components switched on. The results for one tune of the old model (DW)
and one of the new (Perugia 0) are shown in
figs.~\ref{fig:b_corr_DW_parts} and \ref{fig:b_corr_P0_parts}, respectively. 
In both cases, and also for the other tunes not shown here, the general
trend is for the MPI component to dominate the distributions. 
Again, this has partly to be understood
in the light of the MPI component generating the largest part of the
multiplicity, see table \ref{tab:mult_parts}, such that statistical
fluctuations are relatively more important when that component is
switched off, as in the RAD and HARD samples. 

As discussed above, however, one notes that the HARD component by
itself only produces very short-range correlations, that drop off
quickly to zero. Adding parton showers, in the RAD samples, extends
the reach of these correlations somewhat further in $\eta$, including
a small tail towards very large $\eta$, presumably generated by
initial-state radiation from the beams.

Interestingly, the behaviour of the MPI component is
somewhat different between the two kinds of models. In the old model,
fig.~\ref{fig:b_corr_DW_parts}, 
the MPI component becomes completely dominant at large $\eta$ and
there has the same magnitude as the low-\pt\ sample itself. In the new
model, fig.~\ref{fig:b_corr_P0_parts}, 
the MPI component alone drops off and is eclipsed by
the shower component at the highest values of $\eta$, indicating a
qualitative difference between the models, consistent with
the new model deriving more of its total particle production from
shower-related activity. 
\label{sec:b-procs}

\begin{figure}[t]
  \centering
  \includegraphics[width=0.50\textwidth]{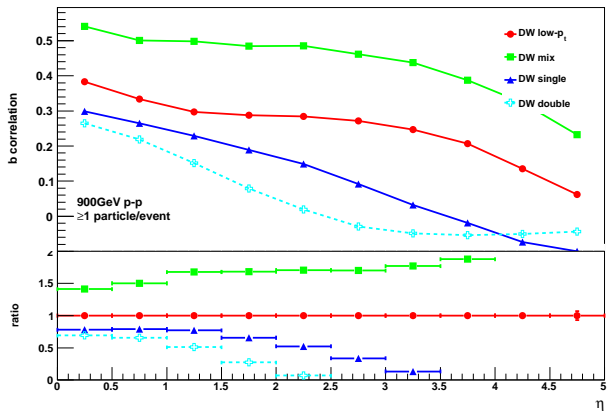}
  \caption{Inclusive $b$ correlation distribution for tune DW minimum
    bias sub-processes. Lower pane: ratio to the
    distribution of the low-\pt sample.}
  \label{fig:b_corr_DW_procs}
\end{figure}

The effects of diffractive components, a non-zero contamination of
which may be present especially in very  inclusive minimum-bias
measurements, are illustrated in fig.\ \ref{fig:b_corr_DW_procs}. 
The correlations in the SD and DD samples are
intrinsically shorter-range than those of their non-diffractive
counterparts, consistent with diffractive systems having a limited
extension in rapidity. 

However, we also see an interesting effect of combining
the samples, namely that the correlations in the combined sample are
\emph{stronger} than in any of the individual components. It seems
that, by mixing in less correlated diffractive components, we have
actually enhanced the final correlations. What is going on? This
effect can be illustrated by imagining we have two separate
distributions, $A$ and $B$ (in our case represented, e.g., by
the diffractive and non-diffractive samples). Imagine further that
the fluctuations inside each sample are purely statistical, for
illustration, such that
the correlation strength inside each sample is zero.
What will happen when we look at the combination $A+B$? If the mean of
$A$ is smaller than that of $B$, 
then \emph{every} event of type $A$ will look like it fluctuated down,
systematically, from the mean of $A+B$, and conversely for the $B$
sample. In their combination, therefore, we will see a non-zero
correlation if the mean values are different. Since the
diffractive and non-diffractive event samples have very different
average multiplicities (see table~\ref{tab:mult_procs}) this effect
will lead to an increase in the correlations in the combined sample, as
observed in fig.~\ref{fig:b_corr_DW_procs}. 

\subsection{`Twisted' b Correlations \label{sec:b-twisted}}

We now turn to the dependence on azimuth of the forward-backward
correlation strengths. A related type of correlations sensitive to
both $\eta$ and azimuthal $\phi$ were recently highlighted by the CMS
experiment \cite{Khachatryan:2010gv} and have stimulated quite a lot
of interest due to  the observation of the so-called ``ridge effect''
in high-multiplicity events. The correlations presented in this report
are somewhat simpler in spirit, and our focus is not
primarily on high multiplicities, but 
we note that it could be an interesting follow-up
study to determine whether twisted $b$-correlations
could also be used to shed more light on the ridge. 

We shall study the $\phi$ dependence of the $b$-correlations in two
ways. The first is based only on the detector geometry. As this is 
independent of the event shape, no preference is given
to any particular direction. The second method gives preference to  
the direction of the leading charged particle in the event. This will
bias the zero point in $\phi$ to coincide with the most active part of
the event.   

In each case, we divide the $\phi$-plane into three regions of size 
$\Delta\phi = 2/3\pi$. For the detector-defined geometry, we define a
\emph{parallel}, an \emph{opposite}, and a \emph{transverse}
region. (Note: we use ``parallel'' and ``opposite'' here, to
distinguish the geometry from Field's ``towards'' and ``away''
regions, the latter of which we take to be defined relative to the direction
of a lead particle or jet and not by the absolute detector geometry.) 
Quite arbitrarily, we define 
\begin{itemize}
\item A \emph{parallel} region covering the region 
$-\pi < \phi < -1/3\pi$ in absolute azimuth, 
\item An \emph{opposite} region covering $ 0 < \phi < 2/3\pi$, 
\item A \emph{transverse} region occupying the region between these,
  i.e., the slices  
$-1/3\pi < \phi < 0$ and  $2/3\pi < \phi < \pi$. 
\end{itemize}
In calculating the
correlation between $\eta-\phi$ regions the comparison is always to
the \emph{parallel} case on one side. We are aware that this is quite crude
and that one could increase statistics by integrating over the
location of the arbitrary azimuthal zero point, but point out that this is
intended merely as a first exploration of 
the properties of `twisted' correlations. 

The terms of the $b$ correlation expression now refer to $\eta$-bins with a 
$\phi$-dependence. Hence, the correlation expression must include this new 
degree of freedom. Since all regions are a priori equivalent, 
the normalizing terms in $b$, $\left<n_{f}\right>^{2}$ and
$\left<n_{f}^{2}\right>$, are taken simply from the parallel one. 
Only the product of activity in corresponding bins of $\eta-\phi$  
are sensitive to the variation in $\phi$ region. The new expression,
$b_\phi^{\mr{twist}}$, for the correlation becomes: 
\begin{equation}
  \label{eqn:b-twist}
  \centering
  b_\phi^{\mr{twist}} =  \frac{\left<n_{b,\phi}n_{f,\parallel}\right> - \left<n_{f,\parallel}\right>^{2}}{\left<n_{f,\parallel}^{2}\right> - \left<n_{f,\parallel}\right>^{2}} .
\end{equation}

\begin{figure}[t]
  \centering
  \includegraphics[width=0.50\textwidth]{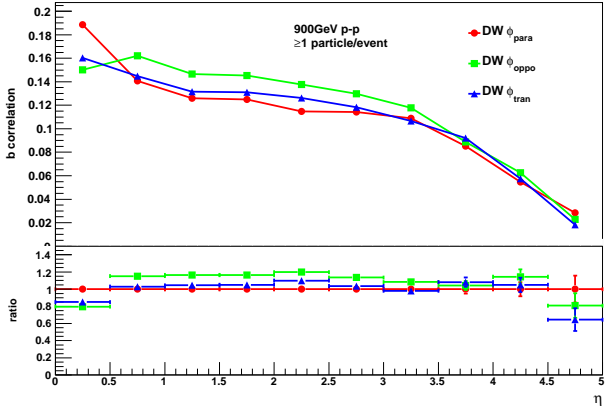}
  \caption{$b^{\mr{twist}}_\phi$ correlation distributions for hadron-level charged particles for DW Pythia tune for the three different combinations
    of $\phi$ regions, defined with respect to the absolute 
    detector geometry. Lower pane: ratio  to the \emph{parallel} distribution.
  \label{fig:DW_phi}}
\end{figure}

\begin{figure}[t]
  \centering
  \includegraphics[width=0.50\textwidth]{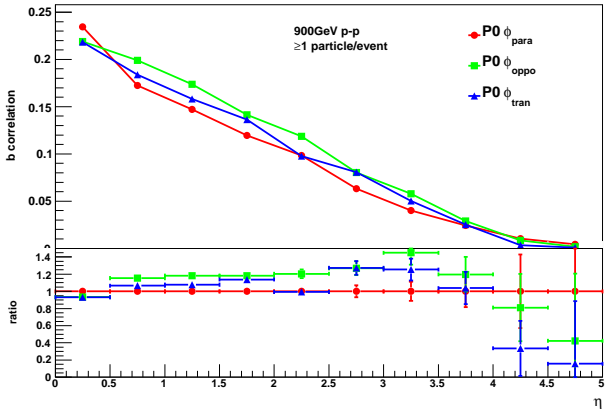}
  \caption{$b^{\mr{twist}}_\phi$ correlation distributions for hadron-level charged
    particles for the Perugia 0 tune for the three different combinations
    of $\phi$ regions, defined with respect to the absolute 
    detector geometry. Lower pane: ratio  to the \emph{parallel}
    distribution.
  \label{fig:P0_phi}}
\end{figure}

Figures \ref{fig:DW_phi} and \ref{fig:P0_phi} show the three different
types of correlations we can obtain for the detector-based geometry, 
for \py\ tunes DW and Perugia 0, respectively. As expected, both
models exhibit a peak in the correlation at low $\eta$ in the \emph{parallel}
region, illustrating that the low-$\eta$ correlation is also most
pronounced at low $\Delta\phi$. Beyond the first bin in $\eta$,
however, the \emph{opposite} correlation is strongest. This follows
from momentum conservation, smeared out over
$\eta$. The correlation with the \emph{transverse} region is as close
as we can come to defining 
an ``underlying event'' in an otherwise featureless minimum-bias
event without a reference direction. 

The difference in correlation strength between the three regions are
not extremely large in absolute terms, however.
This leads us to consider whether there is a way to enhance the
differences while remaining in a minimum-bias context. 
By choosing the zero point of the $\phi$ coordinate, event by
event, to be the direction of the leading (hardest) 
charged particle, we can now use
the nomenclature of Field and define the
\emph{towards} region to include the azimuthal angles $\phi < \pm 1/3\pi$  
around the lead particle, the \emph{away} 
region covers $ \pm2/3\pi < \phi < \pm\pi$ and the \emph{transverse} 
region lies in between, over $\pm1/3\pi < \phi < \pm2/3\pi$. We label
the corresponding correlation $b^{\mr{lead}}_\phi$, to distinguish it
from the detector-based one. 

This orientation has a significant effect in particular for events with 
a semi-hard perturbative scattering, in which the main axis 
in the {transverse} plane becomes oriented to the production axis of the
outgoing partons. The bias towards $\phi=0$ as the direction of the
lead particle means that the three different $\phi$ regions can no longer be
expected to have the same averages and variances. Nonetheless, 
in order to define a measure comparable to the one above, we shall still define
the normalizing terms with respect to the towards region, so
that eq.~(\ref{eqn:b-twist}) still holds, although its statistical
interpretation is modified. 

\begin{figure}[t]
  \centering
  \includegraphics[width=0.50\textwidth]{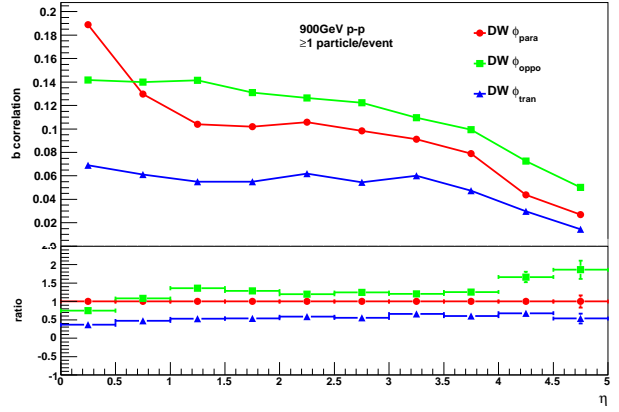}
  \caption{$b^{\mr{lead}}_\phi$ correlation distributions for hadron-level charged
    particles for the DW tune for the three different combinations
    of $\phi$ regions, defined with respect to the lead particle
    trajectory. Lower pane:
    ratio to the \emph{towards} distribution.} 
  \label{fig:DW_lead}
\end{figure}

\begin{figure}[t]
  \centering
  \includegraphics[width=0.50\textwidth]{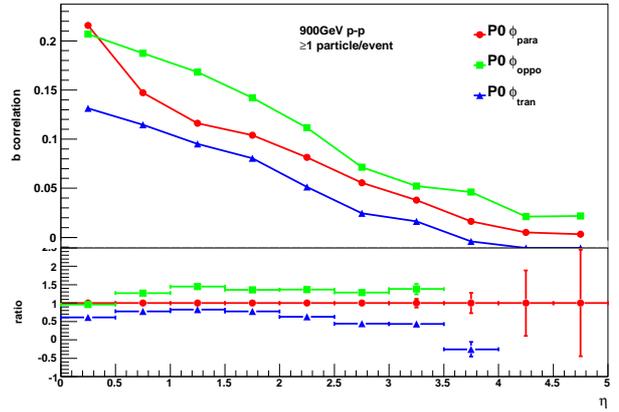}
  \caption{$b^{\mr{lead}}_\phi$ correlation distributions for
    hadron-level charged particles for the Perugia 0 tune 
    for the three different combinations
    of $\phi$ regions, defined with respect to the lead particle
    trajectory. Lower pane:
    ratio  to the \emph{towards} distribution.} 
  \label{fig:P0_lead}
\end{figure}
Redoing the twisted correlations with this definition of the zero
point in azimuth, we obtain figures
\ref{fig:DW_lead} and \ref{fig:P0_lead}, for DW and Perugia 0,
respectively, with the other tunes exhibiting  similar qualitative
features. The general remarks are similar to those for the
detector-based geometry, but the differences between the regions are
much more clearly visible. Also, the \emph{transverse} region can here be clearly
identified as lower than the others, consistent with its being an
``underlying event'' to the production of a ``hard particle''.

\section{Event Shapes \label{sec:shapes}}

A further characteristic of the structure of the events in the
minimum-bias sample can be gained by considering their eigenvalues
along the principal event axes in the transverse plane. Obviously, 
we expect minimum-bias events to be much more uniform than,
e.g., jet events, and hence to have smaller eigenvalues, but
\emph{how} much smaller? What is their average shape, and how much
does it fluctuate between events? 

To address these questions, and to gain a first idea of their
sensitivity to the physics modeling, we consider the transverse thrust
($T_{t}$) and transverse minor ($M_{t}$) values and axes
\cite{Banfi:2010xy}.  

\subsection{Transverse Thrust}

The transverse thrust axis can be found by maximizing the coincidence of an arbitrary vector with the dominant 
direction of particle flow in an event in $\phi$. The value of
transverse thrust  is then defined as:

\begin{equation}
  \label{eqn:thrust}
  \centering
  T_{\perp} = \max_{|\vec{n}_{\perp}|=1} \frac{\sum_{i}|\vec{n}_{\perp}\cdot \vec{p}^{i}_{\perp}|}{\sum_{i}|\vec{p}^{i}_{\perp}|}~,
\end{equation}
where $i$ runs over the charged tracks in the event, $\vec{n}_{\perp}$
is the transverse thrust axis unit vector and  
$\vec{p}^{i}_{\perp}$ is the track transverse momentum vector. The
observable is bounded by $0.5 < T_{\perp} < 1.0$. Dijet-like
events, where the highest-momentum particles are produced
back to back, have a pen\-cil-like 
shape in $\phi$, with particle production aligned predominantly along
the axis. Such events have high transverse thrust values $\sim 1$. 
In contrast, in events where non-pertur\-bative and/or MPI production is
predominant, more particles will lie off the main production axis, giving a more  
circular distribution of tracks in $\phi$, for which the transverse
thrust value will lie closer to $0.5$.

\begin{figure}[t]
  \centering
    \includegraphics[width=0.50\textwidth]{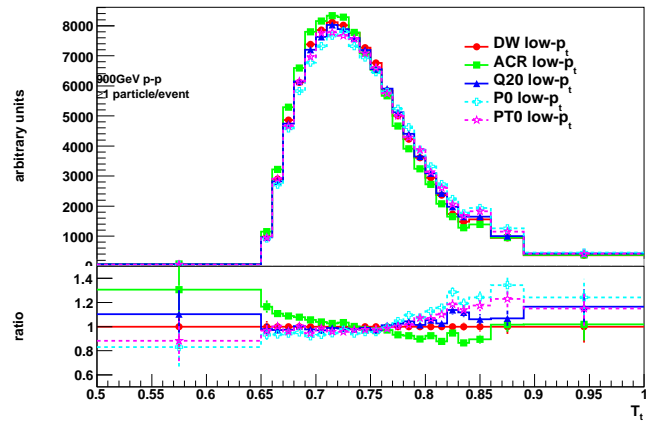}
\caption{Transverse thrust distributions for the low-\pt\ sub-sample
  of selected tunes, normalized to unit area. Lower
  pane: ratio to the 
  DW distribution. } 
  \label{fig:thrust_all}
\end{figure} 
Fig.\ref{fig:thrust_all} shows the transverse thrust distributions of
the low-\pt\ sub-samples of the selected tunes. Somewhat surprisingly,
perhaps, the models agree to within 10--20\% over most of the
range. This presumably reflects the fundamental similarity between the
MPI-based perturbative modeling in these tunes. A comparison with
experimental data could give valuable insights on whether the real
world is more or less ``jetty'' than these model predictions
indicate. 

\begin{figure}[t]
  \centering
  \subfigure[]{
    \includegraphics[width=0.50\textwidth]{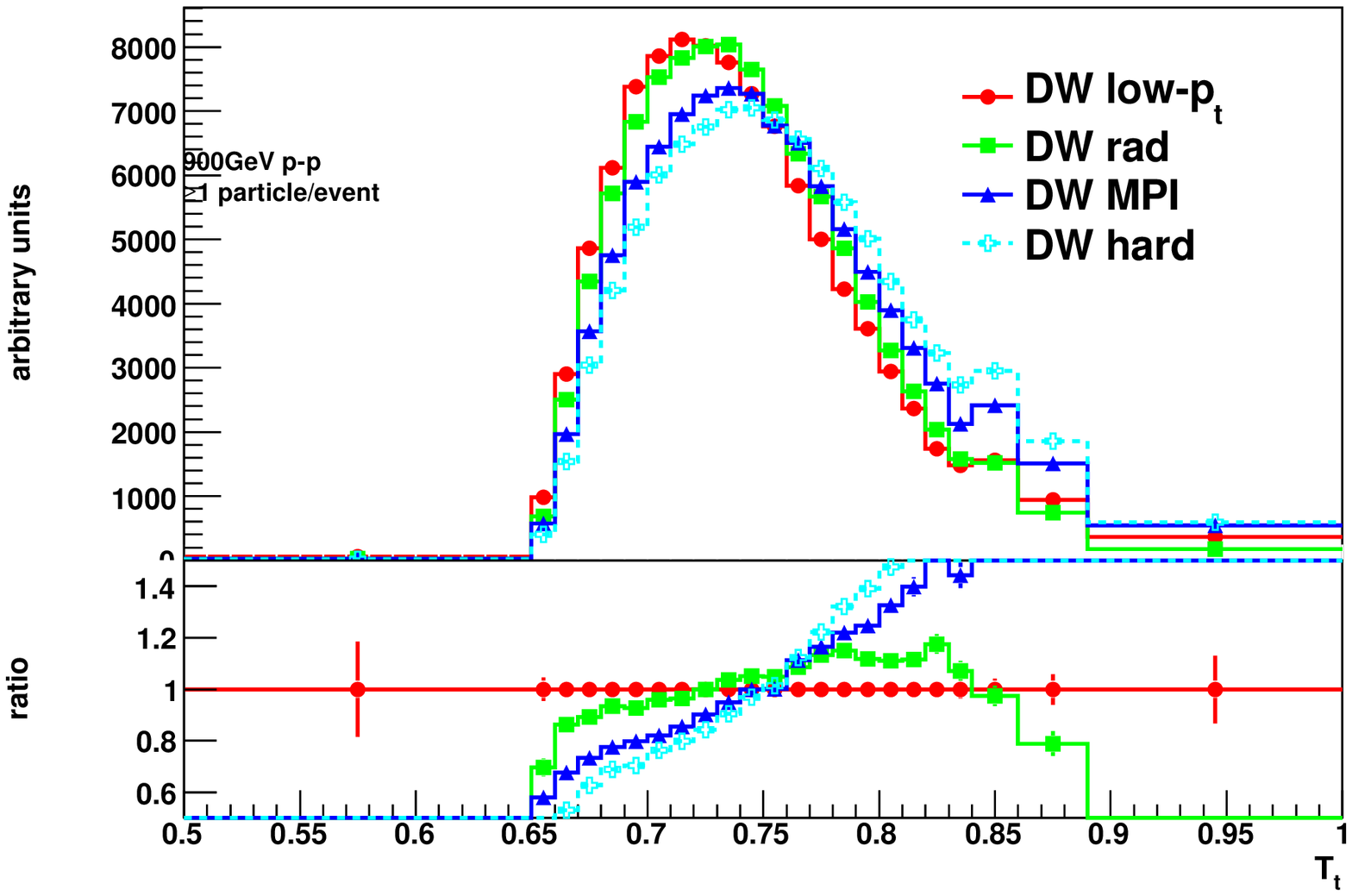}
    }
  \subfigure[]{
    \includegraphics[width=0.50\textwidth]{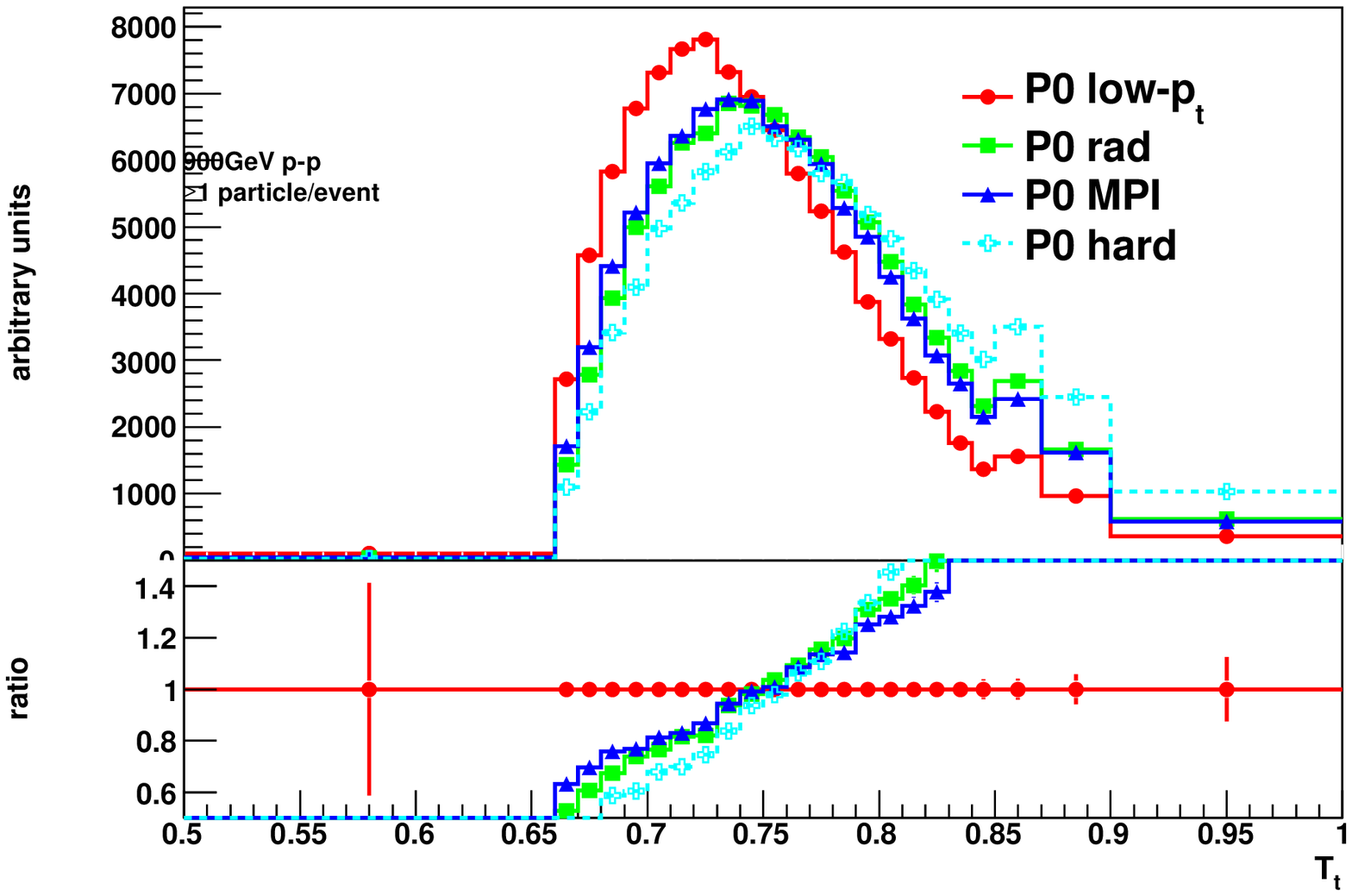}
    }
\caption{Transverse thrust distributions for the low-\pt\ sample for
  a) DW and b) Perugia 0, normalized to unit area. Lower panes: ratio
  to the respective low-\pt samples.}
  \label{fig:thrust_tunes}
\end{figure} 
To further analyze the structure of this distribution,
fig.\ref{fig:thrust_tunes} shows the transverse thrust distributions
for the HARD, RAD, and MPI samples, compared to the full
low-\pt\ simulation, for the DW (above) and Perugia 0 (below)
tunes. For the HARD sample (i.e., before showering and MPI), 
the distributions are more pencil-like, peaking at
somewhat higher values of $T_\perp$, illustrated by the dashed
(cyan) curves. It is interesting that, in the old model (DW), the MPI
component by itself (solid blue lines with triangular symbols) 
only reduces the peak value very slightly, whereas
the addition of radiation (RAD: solid green lines with
square symbols) produces a much larger shift. In the new model,
however, the MPI and RAD samples each appear to give a similar-size
shift. Despite their apparent similarities, there are therefore
still interesting differences underlying these distributions, which,
as we have argued, the measurement of $b$ correlations can help
resolve. 

\subsection{Transverse Minor}

\begin{figure}[t]
  \centering
  \subfigure[]{
    \includegraphics[width=0.50\textwidth]{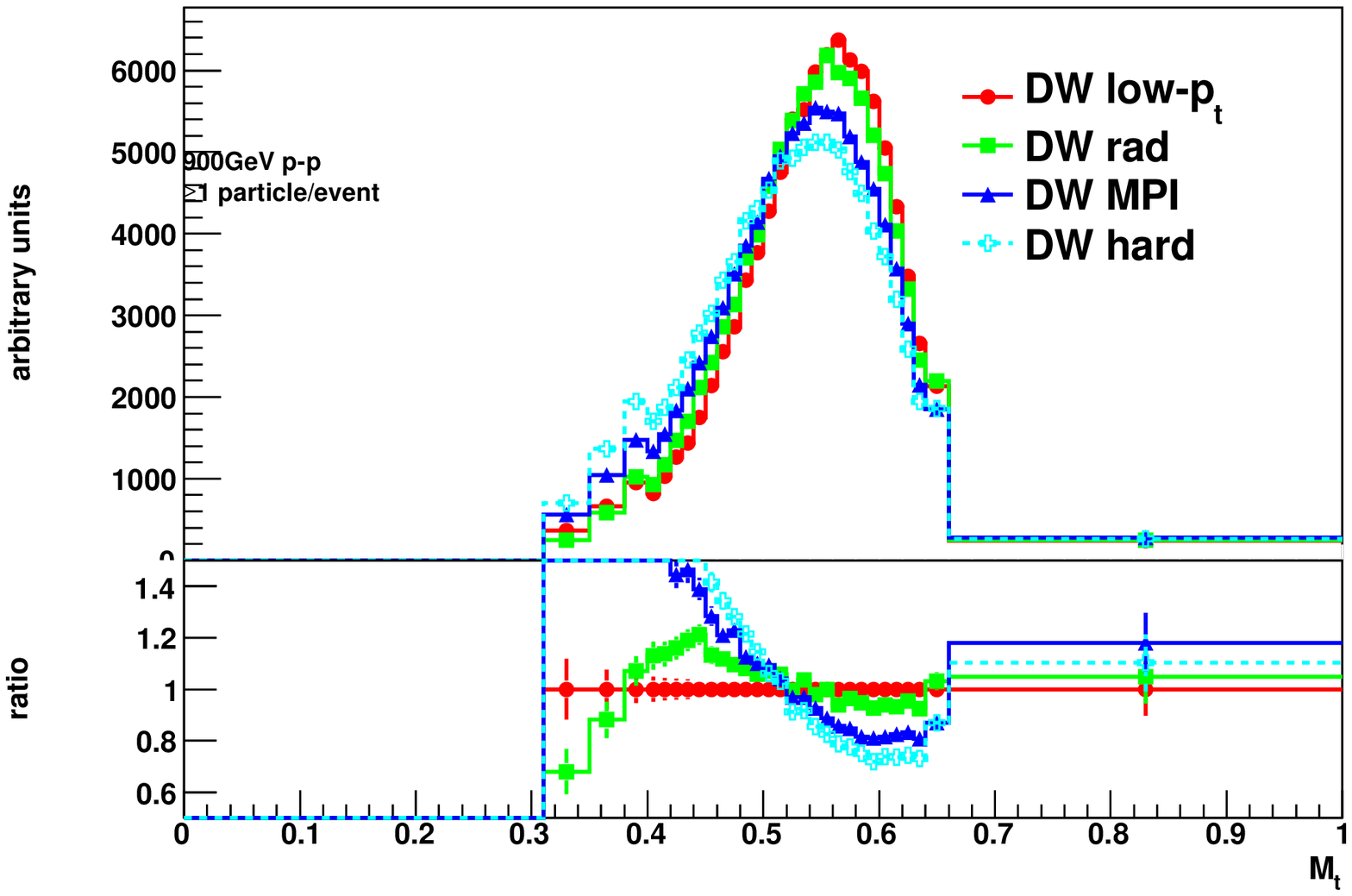}
    }
  \subfigure[]{
    \includegraphics[width=0.50\textwidth]{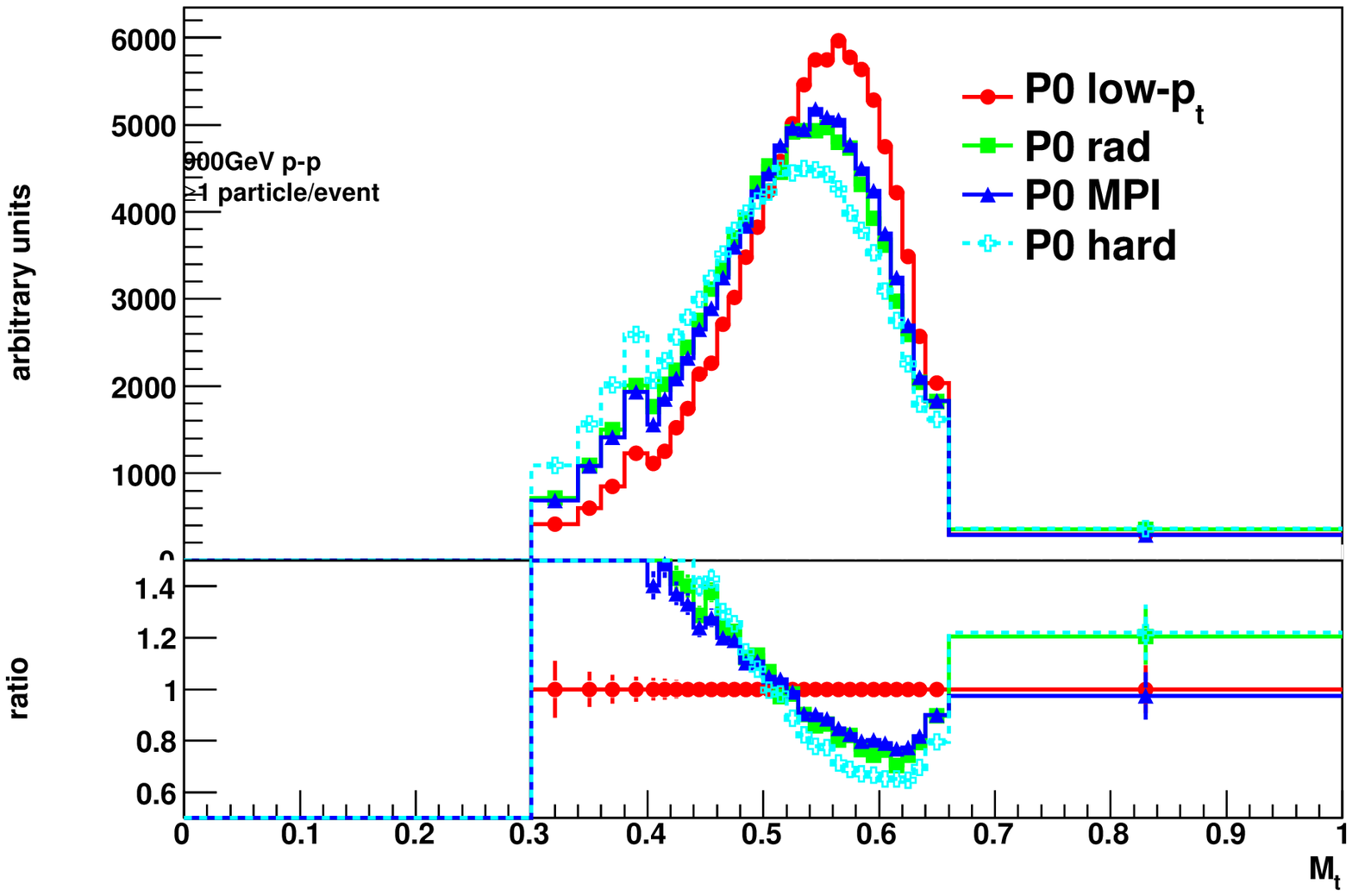}
    }
\caption{Transverse minor distributions for the low-\pt\ sample for a)
  DW and b) Perugia 0, normalized to unit area. Lower panes: ratio
  to the respective low-\pt samples.}
  \label{fig:minor_tunes}
\end{figure}

The transverse minor axis lies perpendicular to the thrust axis in
$\phi$. It is defined as: 

\begin{equation}
  \label{eqn:minor}
  \centering
  M_{\perp} = \max_{|\vec{n}_{\perp}|=1} \frac{\sum_{i}|\vec{n}_{\perp}\times \vec{p}^{i}_{\perp}|}{\sum_{i}|\vec{p}^{i}_{\perp}|}~,
\end{equation}
using the same definitions as in eq.~(\ref{eqn:thrust}). 
\begin{figure}[t]
  \centering
    \includegraphics[width=0.50\textwidth]{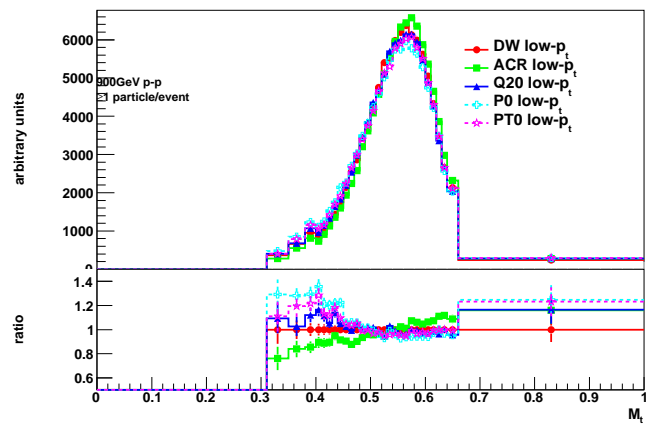}
\caption{Transverse minor distributions for low-\pt\ sub-samples of
  selected tunes, normalized to unit area. Lower pane: 
ratio to the DW distributions. }
  \label{fig:minor_all}
\end{figure} 
Fig.\ref{fig:thrust_all} shows the transverse minor distributions of
the low-\pt\ sub-samples of the selected tunes. As before, the
variations between models is relatively mild, with the contributions
from each model component, fig.~\ref{fig:minor_tunes}, 
exhibiting similar differences as for $T_\perp$.

\section{Conclusion \label{sec:conclusions}}

We have illustrated that forward-backward correlations can be used to
extract information on the relative strengths of different sources of 
particle production in minimum-bias events: models dominated by a
single hard (dijet) interaction exhibit strong short-range
correlations and weak long-range ones, while models with a larger
component of soft production between the remnants generate stronger long-range
correlations.

We propose to add these distributions to the ``standard'' ones already
measured by the LHC experiments, and further to add correlations
between different regions in azimuthal $\phi$, which we label
`twisted' forward-backward correlations. 

We have illustrated these inferences by comparing a small set of
recent tunes of the \py~6 Monte Carlo model. Although they are all
based on a picture of multiple parton interactions (MPI) interfaced to
the Lund string fragmentation model, they differ qualitatively in the
shower and remnant modeling, and quantitatively in the fragmentation
tuning and amount of showering vs.~MPI. 

We further believe that measurements of event shapes, such as transverse
thrust and transverse minor, can help shed light on the overall
properties and structure of minimum-bias events. For instance, a model
with a strong dominance of perturbative (mini-)jet production should
also predict event shapes closer to equivalent pQCD ones in dijet
events, while models characterized by other particle production
mechanisms should exhibit spectra further from the factorized 
pQCD prediction. In that context, however, the models studied here
appear to give relatively similar results, presumably owing to the
significant properties they share at the level of the underlying
perturbative modeling. 

\subsection*{Acknowledgments}
The authors are grateful to S.~Ferrag and to C.~Buttar for their 
help and guidance. 
This research project has been supported by a Marie Curie Early Stage
Research Training Studentship of the European Community's Sixth
Framework Programme under contract number
(MRTN-CT-2006-035606-MCnet).

\bibliography{main-epj}

\end{document}